\documentclass[
]{jss}

\usepackage[utf8]{inputenc}

\author{
Christian Thiele\\University of Applied Sciences Bielefeld \And Gerrit Hirschfeld\\University of Applied Sciences Bielefeld
}
\title{\pkg{cutpointr}: Improved Estimation and Validation of Optimal Cutpoints
in \proglang{R}}

\Plainauthor{Christian Thiele, Gerrit Hirschfeld}
\Plaintitle{cutpointr: Improved estimation and validation of optimal cutpoints in R}
\Shorttitle{\pkg{cutpointr}: Improved Estimation and Validation of Optimal Cutpoints
in \proglang{R}}

\Abstract{
``Optimal cutpoints'' for binary classification tasks are often
established by testing which cutpoint yields the best discrimination,
for example the Youden index, in a specific sample. This results in
``optimal'' cutpoints that are highly variable and systematically
overestimate the out-of-sample performance. To address these concerns,
the \pkg{cutpointr} package offers robust methods for estimating optimal
cutpoints and the out-of-sample performance. The robust methods include
bootstrapping and smoothing based on kernel estimation, generalized
additive models, smoothing splines, and local regression. These methods
can be applied to a wide range of binary-classification and cost-based
metrics. \pkg{cutpointr} also provides mechanisms to utilize
user-defined metrics and estimation methods. The package has
capabilities for parallelization of the bootstrapping, including
reproducible random number generation. Furthermore, it is pipe-friendly,
for example for compatibility with functions from \pkg{tidyverse}.
Various functions for plotting receiver operating characteristic curves,
precision recall graphs, bootstrap results and other representations of
the data are included. The package contains example data from a study on
psychological characteristics and suicide attempts suitable for applying
binary classification algorithms.
}

\Plainkeywords{optimal cutpoint, ROC curve, bootstrap, R}

\Address{
    Christian Thiele\\
  University of Applied Sciences Bielefeld\\
  Interaktion 1 33619 Bielefeld, Germany\\
  E-mail: \email{christian.thiele@fh-bielefeld.de}\\
  URL: \url{http://www.fh-bielefeld.de}\\~\\
    }

\usepackage{amsmath} \usepackage{multirow} \usepackage{booktabs} \usepackage{tabularx} \usepackage{float} \usepackage{caption} 

\begin{document}

\hypertarget{introduction}{%
\section{Introduction}\label{introduction}}

In many applied fields ``optimal cutpoints'' are used to aid the
interpretation of continuous test results or measurements. In medicine,
optimal cutpoints are used to determine the level of a laboratory marker
at which specific interventions should be triggered. In psychology,
optimal cutpoints are used to screen for disorders, for example for
depression or suicidality. Given the importance of cutpoints, a lot of
research is focused on empirically determining ``optimal cutpoints''.
Usually, applied researchers collect data both on the continuous measure
for which a cutpoint is being developed and on a reference test (gold
standard). An optimal cutpoint is then determined by computing a measure
of discriminative ability at all cutpoints and choosing the cutpoint as
``optimal'' that optimizes this measure in the sample.

Unfortunately, cutpoints that are determined this way are highly
sample-specific and overestimate the diagnostic utility of the measure.
Altman et al.~observed more than twenty years ago that different studies
which empirically determined optimal cutpoints for prognostic markers in
breast cancer \citep{altman_dangers_1994} yield highly conflicting
optimal cutpoints. Furthermore, the post-hoc choice of optimal cutpoints
introduces a positive bias to the estimated discriminative ability
\citep{ewald_post_2006, hirschfeld_simulation_2014}. To overcome the
former problem, more robust methods to estimate optimal cutpoints have
been developed \citep{fluss_estimation_2005, leeflang_bias_2008}. In
parallel, methods to estimate the out-of-sample performance of
predictive models have been developed but have not been applied to the
estimation of cutpoint performance. \pkg{cutpointr} is an \proglang{R}
\citep{r_core_team_r:_2018} package that was developed to make robust
estimation procedures for optimal cutpoints and their out-of-sample
performance more accessible (see Section 2 and Section 3). It is
available from the Comprehensive R Archive Network (CRAN) at
\url{https://cran.r-project.org/package=cutpointr}.

Furthermore, \pkg{cutpointr} was developed with three design goals in
mind. First, to make computationally demanding methods such as
bootstrapping feasible, we wanted to improve the performance and
scalability of optimal cutpoint procedures. With larger data (say, more
than 1 million observations) some of the existing solutions and packages
are prohibitively slow or not memory-efficient enough. The
\pkg{cutpointr} package aims at better performance with larger data. The
included bootstrapping routine can easily be run in parallel by setting
the \texttt{allowParallel} parameter to \texttt{TRUE} and registering a
parallel backend before running \texttt{cutpointr}. Second, for
convenient integration into workflows that employ functions from the
``tidyverse'' \pkg{cutpointr} is pipe-friendly by making \texttt{data}
the first argument and providing the output as a data frame which can be
easily piped to further functions. The output is tidy
\citep{wickham2014tidy} in the sense that every column represents a
variable and every observation is a row. It may on the other hand be
untidy in the sense that it contains some columns that are observations
on the level of the model or data set and thus constant over subgroups
(e.g., \texttt{AUC}), if subgroups are specified. The original data, the
receiver operating characteristic (ROC) curve and, if desired, the
results of the bootstrapping are returned in nested tibbles. Third,
\pkg{cutpointr} tries to offer an intuitive interface. For example, it
allows for control over the positive and negative classes as well as the
direction that defines whether higher or lower values of the predictor
stand for the positive class. If these are left out, \pkg{cutpointr}
will attempt to determine these parameters automatically and use
sensible defaults. The package should be applicable in a wide variety of
tasks and tries to avoid focusing on a specific field of study, for
example by a wording of the documentation and function arguments that is
as general as possible.

At present, several other packages for the \proglang{R} environment for
statistical computing and graphics exist that calculate optimal
cutpoints or ROC curves. Most notably, \pkg{OptimalCutpoints}
\citep{lopez-raton_optimalcutpoints:_2014} was specifically designed to
calculate cutpoints. While this package implements a wide range of
metrics for diagnostic utility, it lacks the ability for robust cutpoint
estimation and out-of-sample validation. Additionally, \pkg{GsymPoint}
\citep{lopez-raton_gsympoint:_2017} provides alternative estimation
procedures for the generalized symmetry point and \pkg{ThresholdROC}
\citep{perez-jaume_thresholdroc:_2017} offers several optimization
methods for application on two- and three-class problems. \pkg{ROCR}
\citep{sing_rocr:_2005}, that was developed for ROC curve analysis, can
also be used to calculate optimal cutpoints after writing short
functions that search for the optimal metric value and return the
corresponding cutpoint. \pkg{pROC} \citep{robin_proc:_2011} offers
cutpoint estimation in terms of two metrics including weighting of false
positives and false negatives. Most of these packages do not offer
functions that perform robust estimation of optimal cutpoints or provide
procedures to assess the variability and out-of-sample performance of
the optimal cutpoints.

Furthermore, the packages differ widely in their choice of and control
over the specifics of the cutpoint calculation, such as the direction
and the use of midpoints (table 1). Only some packages offer the choice
of the direction, that is whether higher or lower values of the
predictor stand for the positive or negative class, or in other words
the direction of the ROC curve. Some packages allow for the automatic
detection of a plausible direction. All of the packages except for
\pkg{cutpointr} make the choice of whether or not to use midpoints
implicitly. These seemingly minute choices become relevant when small
datasets are analyzed (simulation results can be found in the package
vignette, see \texttt{vignette("cutpointr")}). \pkg{cutpointr} supports
either choosing the direction of the ROC curve manually or detecting it
automatically. Whether or not midpoints are returned can be specified.

\begin{table}[!htbp]
\begin{tabular}{lllllll}
\hline
  & \multirow{2}{*}{Cutpoint}  & \multirow{2}{*}{Robust}  &   \multicolumn{3}{c}{Direction}     & \multirow{2}{*}{Returns} \\
\cline{4-6}
Package   & \multirow{2}{*}{optimization} & \multirow{2}{*}{cutpoints}  & \multirow{2}{*}{Settings}  & \multirow{2}{*}{Behavior}  & Automatic  & \multirow{2}{*}{midpoints} \\
          &                               &                             &                            &                            & available  &  \\
\hline
\multirow{2}{*}{cutpointr}  & \multirow{2}{*}{yes} & \multirow{2}{*}{yes} & >=  & >= & \multirow{2}{*}{yes} & \multirow{2}{*}{selectable} \\
\vspace{0.3cm}
       &       &    & <=         & <=       &            &  \\
\multirow{2}{*}{OptimalCutpoints} & \multirow{2}{*}{yes} & \multirow{2}{*}{no} & >  & > & \multirow{2}{*}{no} & \multirow{2}{*}{no} \\
\vspace{0.3cm}
                            &   &    & <          & <        &            &  \\
\multirow{2}{*}{pROC} & \multirow{2}{*}{yes} & \multirow{2}{*}{no} & > & > & \multirow{2}{*}{yes} & \multirow{2}{*}{yes} \\
\vspace{0.3cm}
                           &    &    & <          & <    &            &       \\
\vspace{0.3cm}
ROCR                            & no & no & none       & >        & no         & no      \\
\multirow{2}{*}{ThresholdROC} & \multirow{2}{*}{yes} & \multirow{2}{*}{yes} & \multirow{2}{*}{none} & >= & \multirow{2}{*}{yes} & \multirow{2}{*}{no} \\
                           &    &    &           & <=     &            &       \\
\hline
\end{tabular}
\centering
\caption{Features, settings and actual behavior regarding direction and midpoints of relevant packages for ROC curve analysis or cutpoint optimization.}
\end{table}

The rest of the article is structured as follows. In section 2 we
describe the different methods to determine optimal cutpoints, focusing
on both the metrics used to determine cutpoints, for example the Youden
index or cost based methods, as well as the methods to select a
cutpoint, for example LOESS-based smoothing or bootstrapping. In section
3 we describe the bootstrapping method to estimate the out-of sample
performance and the variability of the cutpoints. In section 4 we
compare the performance of different estimation methods on simulated
data. In section 5 we show how these methods are implemented in
\pkg{cutpointr}. In section 6 we show how these can be used to analyze
an example dataset on optimal cutpoints for suicide. In section 7 we
offer some concluding remarks.

\hypertarget{estimating-optimal-cutpoints}{%
\section{Estimating optimal
cutpoints}\label{estimating-optimal-cutpoints}}

Optimal cutpoints are estimated by choosing a specific metric that is
being used to quantify the discriminatory ability and a specific method
to select a cutpoint as optimal based on this metric. The most
widely-used combination of these two is to use the Youden index as a
metric and to select the cutpoint as optimal that empirically maximizes
the Youden index in-sample. In the following, we discuss several
alternative metrics and methods. While the choice of metrics has
received ample attention, the methods to determine optimal cutpoints
have been largely neglected.

\hypertarget{metrics-to-quantify-discriminatory-ability}{%
\subsection{Metrics to quantify discriminatory
ability}\label{metrics-to-quantify-discriminatory-ability}}

Most of the metrics to assess the discriminatory ability of a cutpoint
rely on the elements of the \(2 \times 2\) confusion matrix, that is
true positives (\(TP\)), false positives (\(FP\)), true negatives
(\(TN\)), and false negatives (\(FN\)). Sensitivity is defined as
(\(TP\)) divided by the total number of positives
\(Se = \frac{TP}{TP + FN}\). Specificity is defined as \(TN\) divided by
the total number of negatives \(Sp = \frac{TN}{TN + FP}\). When
searching over different cutpoints, there is a trade-off between \(Se\)
and \(Sp\) and thus the researcher often tries to optimize both of these
measures simultaneously. This can, for example, be achieved by
maximizing the Youden- or J-Index (available in the function
\texttt{youden}), that is \(Se + Sp - 1\). Alternatively, \(Se + Sp\)
can be maximized (available in the function \texttt{sum\_sens\_spec})
which leads to the selection of the same cutpoint. As a way to balance
\(Se\) and \(Sp\) the absolute difference \(|Se - Sp|\) can be minimized
(available in the function \texttt{abs\_d\_sens\_spec}). An alternative
to \(Se\) and \(Sp\) are the positive and negative predictive values.
The positive predictive value is defined as \(PPV = \frac{TP}{TP + FP}\)
and the negative predictive value as \(NPV = \frac{TN}{TN + FN}\).
Again, metrics such as the sum of \(PPV\) and \(NPV\) (available in the
function \texttt{sum\_ppv\_npv}) and the absolute difference between the
two (available in the function \texttt{abs\_d\_ppv\_npv}) can be
optimized.

Sometimes it may be useful to optimize \(Se\) for a desired minimum
value of \(Sp\) or the other way around. For example, to restrict
misclassification of negative cases in a clinical setting, the cutpoint
of a test may be expected to satisfy \(Sp > 0.9\) while maximizing
\(Se\). This type of metric is implemented in \texttt{sens\_constrain}
and, more general and applicable for all metric functions, in
\texttt{metric\_constrain}. These functions will return \(0\) at
cutpoints where the constraint is not met. In a similar vein, different
costs can be assigned to \(FP\) and \(FN\) via
\texttt{misclassification\_cost}.

Furthermore, \pkg{cutpointr} includes some metrics that are common in
document retrieval and some utility metrics, for example \texttt{tp} for
the number of true positives, that are mainly useful in plotting. See
table 2 for a complete overview of included metrics.

Since many metrics rely on \(Se\) and \(Sp\), it can be informative to
inspect the ROC curve, which is defined as the plot of \(1 - Sp\) on the
x-axis and \(Se\) on the y-axis or analogously the false positive rate
on the x-axis and the true positive rate on the y-axis. Usually, this
plot is displayed on the unit square. All points on the ROC curve
correspond to a combination of \(Se\) and \(Sp\) and thus to a
combination of \(TP\), \(FP\), \(TN\) and \(FN\) depending on a
cutpoint. That way, any metric that is calculated from these values can
be calculated for all points on the ROC curve and it is straightforward
to display the chosen cutpoint on the ROC curve. Since the ROC curve is
the visualization of a binary classification depending on possible
cutoff values to achieve that classification, it only depends on the
ranking of the predictions and is insensitive to monotone
transformations of those values.

\begin{table}[]
\centering
\begin{tabularx}{\textwidth}{lX} \\ \hline
Metric                                                         & Description                                                                                        \\ \hline
\texttt{tp}                                                & True Positives: $TP$                                                                                    \\
\texttt{fp}                                               & False Positives: $FP$                                                                                        \\
\texttt{tn}                                                 & True Negatives: $TN$                                                                                      \\
\texttt{fn}                                                & False Negatives: $FN$                                                                                        \\
\texttt{tpr}                                              & True Positive Rate: $TPR$                                                                                     \\
\texttt{fpr}                                             & False Positive Rate: $FPR$                                                                                      \\
\texttt{tnr}                                            & True Negative Rate: $TNR$                                                                                        \\
\texttt{fnr}                                              & False Negative Rate: $FNR$                                                                                      \\
\texttt{plr}                                             & Positive Likelihood Ratio: $PLR = TPR / FPR$                                                                  \\
\texttt{nlr}                                               & Negative Likelihood Ratio: $NLR = FNR / TNR$                                                                  \\
\texttt{accuracy}                                            & Fraction correctly classified: $\frac{TP + TN}{TP + FP + TN + FN}$                                                \\
\texttt{sensitivity} or \texttt{recall}                    & Sensitivity: $Se = TP / (TP + FN)$                                                                            \\
\texttt{specificity}                                         & Specificity: $Sp = TN / (TN + FP)$                                                                            \\
\texttt{sum\_sens\_spec}                                     & Sum of sensitivity and specificity: $Se + Sp$                                                                     \\
\texttt{youden}                                              & Youden- or J-Index: $Se + Sp - 1$                                                                                 \\
\texttt{abs\_d\_sens\_spec}                                  & Absolute difference between sensitivity and specificity: \newline $|Se - Sp|$                                       \\
\texttt{prod\_sens\_spec}                                    & Product of sensitivity and specificity: $Se * Sp$                                                                 \\
\texttt{ppv} or \texttt{precision}                         & Positive Predictive Value: $PPV = TP / (TP + FP)$                                                         \\
\texttt{npv}                                                 & Negative Predictive Value: $NPV = TN / (TN + FN)$                                                             \\
\texttt{sum\_ppv\_npv}                                       & Sum of Positive Predictive Value and Negative Predictive Value: $PPV + NPV$                                       \\
\texttt{abs\_d\_ppv\_npv}                                    & Absolute difference between PPV and NPV: $|PPV - NPV|$                                                            \\
\texttt{prod\_ppv\_npv}                                      & Product of PPV and NPV: $PPV * NPV$                                                                               \\
\texttt{metric\_constrain}                    & Value of a selected metric when constrained by another metric, i.e., $main\_metric = 0$ if $constrain\_metric < min\_constrain$   \\
\texttt{sens\_constrain}                                     & Sensitivity constrained by specificity: \newline $Se = 0$ if $Sp < min\_constrain$                     \\
\texttt{spec\_constrain}                                     & Specificity constrained by sensitivity: \newline $Sp = 0$ if $Se < min\_constrain$                     \\
\texttt{acc\_constrain}                                      & Accuracy constrained by sensitivity: \newline $Acc = 0$ if $Se < min\_constrain$                     \\
\texttt{roc01}                                               & Distance of the ROC curve to the point of perfect discrimination $(0,1)$: \newline $\sqrt{(1 - Se)^2 + (1 - Sp)^2}$ \\
\texttt{F1\_score}                                           & F1-Score: $(2 * TP) / (2 * TP + FP + FN)$                                                                       \\
\texttt{cohens\_kappa}                                       & Cohen's Kappa                                                                                                    \\
\texttt{p\_chisquared}                                       & p-value of a $\chi^2$-test on the confusion matrix                                                               \\
\texttt{odds\_ratio}                                         & Odds Ratio: $\frac{TP / FP}{FN / TN}$                                                                             \\
\texttt{risk\_ratio}                                         & Risk Ratio: $\frac{TP}{TP + FN} / \frac{FP}{FP + TN}$                                                             \\
\texttt{misclassification\_cost}                             & Misclassification Cost: $cost_{FP} * FP + cost_{FN} * FN$                                                         \\
\texttt{total\_utility}                                      & Total Utility: \newline $utility_{TP} * TP + utility_{TN} * TN - cost_{FP} * {FP} - cost_{FN} * FN$               \\
\texttt{false\_omission\_rate}                               & False Omission Rate: $FN / (TN + FN)$                                                                         \\
\texttt{false\_discovery\_rate}                              & False Discovery Rate: $FP / (TP + FP)$                                                                     \\  \hline
\end{tabularx}
\caption{Metrics in cutpointr.}
\end{table}

\hypertarget{methods-to-select-optimal-cutpoints}{%
\subsection{Methods to select optimal
cutpoints}\label{methods-to-select-optimal-cutpoints}}

Most of the methods to estimate optimal cutpoints have been developed
using the Youden index as \texttt{metric}
\citep{fluss_estimation_2005, hirschfeld_simulation_2014, leeflang_bias_2008}.
Regarding these methods and the ones that are included in
\pkg{cutpointr}, the nonparametric empirical method, local regression
(LOESS) smoothing, spline smoothing, generalized additive model (GAM)
smoothing and bootstrapping can be applied to other metrics as well,
while the Normal method and kernel-based smoothing are specific to the
estimation of the Youden index. The former methods rely on optimizing
the function of metric value per cutpoint, denote that function by
\(m(c)\). The nonparametric empirical method picks the cutpoint that
yields the optimal metric value in-sample, the smoothing methods (LOESS,
spline smoothing and GAM) do so after smoothing \(m(c)\).

Since some of the methods do not simply select a cutpoint from the
observed predictor values by optimizing \(m(c)\), it is questionable
which values for \(Se\) and \(Sp\) should be reported. However, since
there exist several methods that do not select cutpoints from the
available observations and in order to unify the reporting of metrics
for these methods, \pkg{cutpointr} reports all standard metrics, for
example \(Se\) and \(Sp\), based on the unsmoothed ROC curve. The main
metric is reported after smoothing and prefixed according to the applied
\texttt{method} function. For example, a method may smooth \(m(c)\)
using a GAM and select a cutpoint that leads to a metric value of
\(Se + Sp = 1.7\) after smoothing, which will be returned in
\texttt{gam\_sum\_sens\_spec}. When looking up the corresponding point
on the unsmoothed ROC curve, that cutpoint may lead to \(Se = 0.8\) and
\(Sp = 0.75\). The latter values will be returned in
\texttt{sensitivity} and \texttt{specificity}. Additional metrics based
on the unsmoothed ROC curve can be added with \texttt{add\_metric}. If
estimates of the variability of the metrics are sought, the user is
referred to the available bootstrapping routine that will be introduced
later. If a method is superior, this should lead to a better in- and
out-of-sample performance in the bootstrap validation.

\hypertarget{nonparametric-empirical-method}{%
\subsubsection{Nonparametric empirical
method}\label{nonparametric-empirical-method}}

The most simple method to optimize a metric is the search over all
cutpoints and picking the cutpoint that yields the optimal metric value
in-sample, optimizing \(m(c)\). This is equivalent to selecting a
cutpoint based on analysis of the (unsmoothed) ROC curve. When applied
to the Youden index, this kind of empirical optimization has been shown
to lead to highly variable results as it is sensitive to randomness in
the sample
\citep{ewald_post_2006, fluss_estimation_2005, hirschfeld_simulation_2014, leeflang_bias_2008}.

\hypertarget{bootstrapping-for-cutpoint-estimation}{%
\subsubsection{Bootstrapping for cutpoint
estimation}\label{bootstrapping-for-cutpoint-estimation}}

Aggregating bootstrapped results of multiple models (``bagging'') can
substantially improve performance of a wide range of models in
regression as well as in classification tasks. In the context of
generating a numerical output, a number of bootstrap samples is drawn
and the final result is the average of all models that were fit to the
bootstrap samples \citep{breiman_bagging_1996}. This principle is
available for cutpoint estimation via the
\texttt{maximize\_boot\_metric} and
\linebreak \texttt{minimize\_boot\_metric} functions. If one of these
functions is used as \texttt{method}, \texttt{boot\_cut} bootstrap
samples are drawn and the nonparametric empirical method is carried out
in each one of them. Finally, a summary function, by default the mean,
is applied to the optimal cutpoints that were obtained in the bootstrap
samples and returned as the optimal cutpoint. If bootstrap validation is
run, i.e., if \texttt{boot\_runs} is larger zero, an outer bootstrap
will be executed, so that \texttt{boot\_runs} bootstrap samples are
generated and each one is again bootstrapped \texttt{boot\_cut} times
for the cutpoint estimation. This may lead to long run times, so
activating the built-in parallelization may be advisable. The advantages
of the bootstrap method are that it doesn't have tuneable parameters,
unlike for example the LOESS smoothing, that it doesn't rely on
assumptions, unlike for example the Normal method, and that it is
applicable to every metric that can be used with
\texttt{minimize\_metric} or \texttt{maximize\_metric}, unlike the
Kernel and Normal methods. Furthermore, the bootstrapped cutpoints
cannot be overfit by running an excessive amount of repetitions
\citep{breiman_random_2001}.

\hypertarget{smoothing-via-generalized-additive-models}{%
\subsubsection{Smoothing via Generalized Additive
Models}\label{smoothing-via-generalized-additive-models}}

The function \(m(c)\) can be smoothed using Generalized Additive Models
(GAM) with smooth terms. This is akin to robust ROC curve fitting in
which a nonparametric curve is fit over the empirical ROC curve.
Subsequently, an estimate of the maximally obtainable sum of \(Se\) and
\(Sp\) can be obtained from that curve. Internally, \texttt{gam} from
the \pkg{mgcv} package carries out the smoothing which can be customized
via the arguments \texttt{formula} and \texttt{optimizer}, see
\texttt{help("gam",\ package\ =\ "mgcv")}. Most importantly, the GAM can
be specified by altering the default formula, for example the smoothing
function \texttt{s} can be configured to utilize cubic regression
splines (\texttt{"cr"}) as the smooth term. The default GAM is of the
form

\[m_i \sim f(c_i) + \epsilon_i\] where \(m\) are the metric values per
cutpoint \(c\), \(f\) is a thin plate regression spline and \(\epsilon\)
is i.i.d. \(N(0, \sigma^2)\) \citep{wood_mgcv:_2001}. An attractive
feature of the GAM smoothing is that the default values tend to work
quite well, see section 4, and usually require no tuning, eliminating
researcher degrees of freedom.

\hypertarget{spline-smoothing}{%
\subsubsection{Spline smoothing}\label{spline-smoothing}}

In the same fashion \(m(c)\) can be smoothed using smoothing splines
which is implemented using \texttt{smooth.spline} from the \pkg{stats}
package in \texttt{maximize\_spline\_metric} and
\linebreak \texttt{minimize\_spline\_metric}. The function to calculate
the number of knots is by default equivalent to the function from
\pkg{stats} but alternatively the number of knots can be set manually
and also the other smoothing parameters of \texttt{stats::smooth.spline}
can be set as desired. For details see
\texttt{?maximize\_spline\_metric}. Spline smoothing finds a cubic
spline \(g\) that minimizes

\[\sum^n_1(y_i - g(x_i))^2 + \lambda \int^{+ \infty}_{-\infty} [g^{''}(z)]^2 dz\]

where the smoothing parameter \(\lambda\) is a fixed tuning constant
\citep{buja_linear_1989}. That makes spline smoothing intuitively
appealing, especially for data with a large number of unique cutpoints,
as the true \(m(c)\) should usually be expected to be quite smooth in
that case and large second derivatives can be directly penalized via
\(\lambda\). However, based on our experience, there is no value of
\(\lambda\) that can be generally recommended and thus it may have to be
tuned, preferably via a validation routine such as cross-validation or
bootstrapping. We assume that smoothing is most attractive for the
aforementioned case of many unique cutpoints in which a large smoothing
parameter can be chosen, which is why in \pkg{cutpointr} the smoothing
parameter \texttt{spar} is by default set to 1. In our simulations,
spline smoothing represents an improvement over the nonparametric
empirical method but seems to be inferior to GAM smoothing or
bootstrapping, see section 4.

\hypertarget{loess-method}{%
\subsubsection{LOESS method}\label{loess-method}}

Additionally, LOESS can be used to smooth \(m(c)\). In simulation
studies this procedure performed favorably in comparison to the
nonparametric empirical method when an automatic LOESS smoothing was
used \citep{leeflang_bias_2008} which is available in the
\texttt{loess.as} function from the \pkg{fANCOVA}
package\citep{wang_fancova:_2010}. Here, the smoothing parameter (span)
is automatically determined using a corrected AIC. This is attractive in
the case of cutpoint selection since it is less susceptible to
undersmoothing than other methods \citep{hurvich_smoothing_1998}.
However, based on our simulations, the LOESS method for cutpoint
selection with smoothing parameter selection based on AICc still tends
to undersmooth \(m(c)\). Importantly, the automatic smoothing is
sensitive to the number of unique cutpoints. Sparse data with a low
number of possible cutpoints is often smoothed appropriately while for
example large samples from a normal distribution are often
undersmoothed. In that scenario the automatically smoothed function
tends to follow \(m(c)\) quite closely which defies the purpose of the
smoothing. Since there is no other rule to apply for the selection of
the optimal smoothing parameters, for example the degree of the
function, this has to be done by hand or by grid search and
cross-validation if the automatic procedure leads to undesirable
results. In \pkg{cutpointr} \(m(c)\) can be smoothed using LOESS and
automatic smoothing parameter selection via the
\texttt{minimize\_loess\_metric} and \texttt{maximize\_loess\_metric}
method functions.

\hypertarget{parametric-method-assuming-normality}{%
\subsubsection{Parametric method assuming
normality}\label{parametric-method-assuming-normality}}

In addition to these general methods, two methods to specifically
estimate the Youden index are included, the parametric Normal method and
a method based on kernel smoothing. The Normal method in
\texttt{oc\_youden\_normal} is a parametric method for maximizing the
Youden index or analogously \(Se + Sp\) \citep{fluss_estimation_2005}.
It relies on the assumption that the predictor for both the negative and
positive observations is normally distributed. In that case it can be
shown that

\[c^* = \frac{(\mu_P \sigma_N^2 - \mu_N \sigma_P^2) - \sigma_N \sigma_P \sqrt{(\mu_N - \mu_P)^2 + (\sigma_N^2 - \sigma_P^2) log(\sigma_N^2 / \sigma_P^2)}}{\sigma_N^2 - \sigma_P^2}\]

where the negative class is normally distributed with
\(\sim N(\mu_N, \sigma_N^2)\) and the positive class independently
normally distributed with \(\sim N(\mu_P, \sigma_P^2)\) provides the
optimal cutpoint \(c^*\). If \(\sigma_N\) and \(\sigma_P\) are equal,
the expression can be simplified to \(c^* = \frac{\mu_N + \mu_P}{2}\).
The \texttt{oc\_youden\_normal} method in \pkg{cutpointr} always assumes
unequal standard deviations.

\newpage

\hypertarget{nonparametric-kernel-method}{%
\subsubsection{Nonparametric Kernel
method}\label{nonparametric-kernel-method}}

A nonparametric method to optimize the Youden index is the Kernel
method. Here, the empirical distribution functions are smoothed using
the Gaussian kernel functions
\[\hat{F}_N(t) = \frac{1}{n} \sum^n_{i=1} \Phi(\frac{t - y_i}{h_y}) \text{ and } \hat{G}_P(t) = \frac{1}{m} \sum^m_{i=1} \Phi(\frac{t - x_i}{h_x})\]
for the negative and positive classes respectively. Following
Silverman's plug-in ``rule of thumb'' the bandwidths are selected as
\(h_y = 0.9 * min\{s_y, iqr_y/1.34\} * n^{-0.2}\) and
\(h_x = 0.9 * min\{s_x, iqr_x/1.34\} * m^{-0.2}\) where \(s\) is the
sample standard deviation and \(iqr\) is the inter quartile range
\citep{fluss_estimation_2005}.

It has been demonstrated that AUC estimation is rather insensitive to
the choice of the bandwidth procedure \citep{faraggi_estimation_2002}.
Indeed, as simulations have shown, the choices of the bandwidth and the
kernel function do not seem to exhibit a considerable influence on the
variance or bias of the estimated cutpoints (results not shown). Thus,
the \texttt{oc\_youden\_kernel} function in \pkg{cutpointr} uses a
Gaussian kernel and the direct plug-in method for selecting the
bandwidths. The kernel smoothing is done via the \texttt{bkde} function
from the \pkg{KernSmooth} package \citep{wand_kernsmooth:_2013}.

Again, there is a way to calculate the Youden index from the results of
this method \citep{fluss_estimation_2005} which is

\[\hat{J} = max_c \{\hat{F}_N(c) - \hat{G}_N(c) \}\]

but as before we prefer to report all metrics based on applying the
cutpoint that was estimated using the Kernel method to the unsmoothed
ROC curve.

\hypertarget{manual-mean-and-median-methods}{%
\subsubsection{Manual, mean and median
methods}\label{manual-mean-and-median-methods}}

Lastly, the functions \texttt{oc\_mean} and \texttt{oc\_median} pick the
mean or median, respectively, as the optimal cutpoint. These functions
are mainly useful, if the performance of these methods in the bootstrap
validation is desired, perhaps as a benchmark. Via these functions, the
estimation of the mean or median can be carried out in every bootstrap
sample to avoid biasing the results by setting the cutpoint to the known
values from the full sample. A fixed cutpoint can be set via the
\texttt{oc\_manual} function.

\hypertarget{bootstrap-estimates-of-performance-and-cutpoint-variability}{%
\section{Bootstrap estimates of performance and cutpoint
variability}\label{bootstrap-estimates-of-performance-and-cutpoint-variability}}

An important aspect to evaluating an optimal cutpoint is estimating its
out-of-sample performance. Popular validation methods include training /
test splits, k-fold cross-validation, and different variations of the
bootstrap \citep{altman_what_2000, arlot_survey_2010}. There have been
numerous attempts at quantifying the bias and variance of validation
methods and to improve existing methods. In general, most validation
methods that rely on data splitting and thus train a model on a subset
of the original data are pessimistically biased, e.g., underestimate
accuracy or overestimate RMSE \citep{dougherty_performance_2010}. Some
variants of the bootstrap try to alleviate that bias, for example the
.632-bootstrap that combines in- and out-of-bag estimations of the
performance \citep{efron_improvements_1997}. In general, the regular (or
Leave-one-out) bootstrap is pessimistically biased. In comparison to
k-fold cross-validation the bootstrap is usually more pessimistically
biased but offers a lower variance \citep{molinaro_prediction_2005}. All
in all, it is a legitimate alternative to cross-validation or other
splitting schemes and may even be regarded as the superior method for
internal validation \citep{steyerberg_internal_2001}. In \pkg{cutpointr}
we prefer to be conservative by using the regular Leave-one-out
bootstrap instead of attempting to correct for the bias. For calculation
of confidence intervals between 1000 and 2000 repeats of the bootstrap
are recommended \citep{carpenter2000sm}.

The bootstrap routine proceeds as follows: A sample from the original
data with the same size as the original data is drawn with replacement.
This represents the bootstrap or ``in-bag'' sample. On average, a
bootstrap sample contains 63.2\% of all observations of the full data
set as some observations are drawn multiple times
\citep{efron_improvements_1997}. The cutpoint estimation is carried out
on the in-bag sample and the obtained cutpoint is applied to the in- and
out-of-bag observations and \(TP\), \(FP\), \(TN\) and \(FN\) are
recorded to obtain in- and out-of-bag estimates of various performance
metrics. The results are returned in the \texttt{boot} column of the
\texttt{cutpointr} object and can be summarized for example via the
\texttt{summary} or \texttt{plot\_metric\_boot} functions. In some
plotting functions \pkg{cutpointr} refers to ``in-sample'' data which is
the complete data set in \texttt{data}, not the in-bag data. Whether a
metric was calculated on the in-bag or out-of-bag data is indicated by
the suffix \texttt{\_b} or \texttt{\_oob} in the data frame of bootstrap
results.

As the cutpoint that was identified as optimal depends on the sample at
hand, a researcher might be interested in a measure of the variability
of this cutpoint
\citep{altman_dangers_1994, hirschfeld_establishing_2015}. Additionally,
as was pointed out, different estimation procedures, which are in the
context of \pkg{cutpointr} comprised of the cutpoint estimation method
and possibly also the chosen metric, are likely to exhibit different
amounts of variability. For example, maximizing the odds ratio is likely
to suffer from a larger variance than maximizing the sum of \(Se\) and
\(Sp\). During bootstrapping, the optimal cutpoints on the in-bag
samples are recorded, returned in the \texttt{boot} column of the
\texttt{cutpointr} object and their distribution can be inspected via
the \texttt{summary} or \texttt{plot\_cut\_boot} functions.

\begin{CodeChunk}
\begin{figure}

{\centering \includegraphics{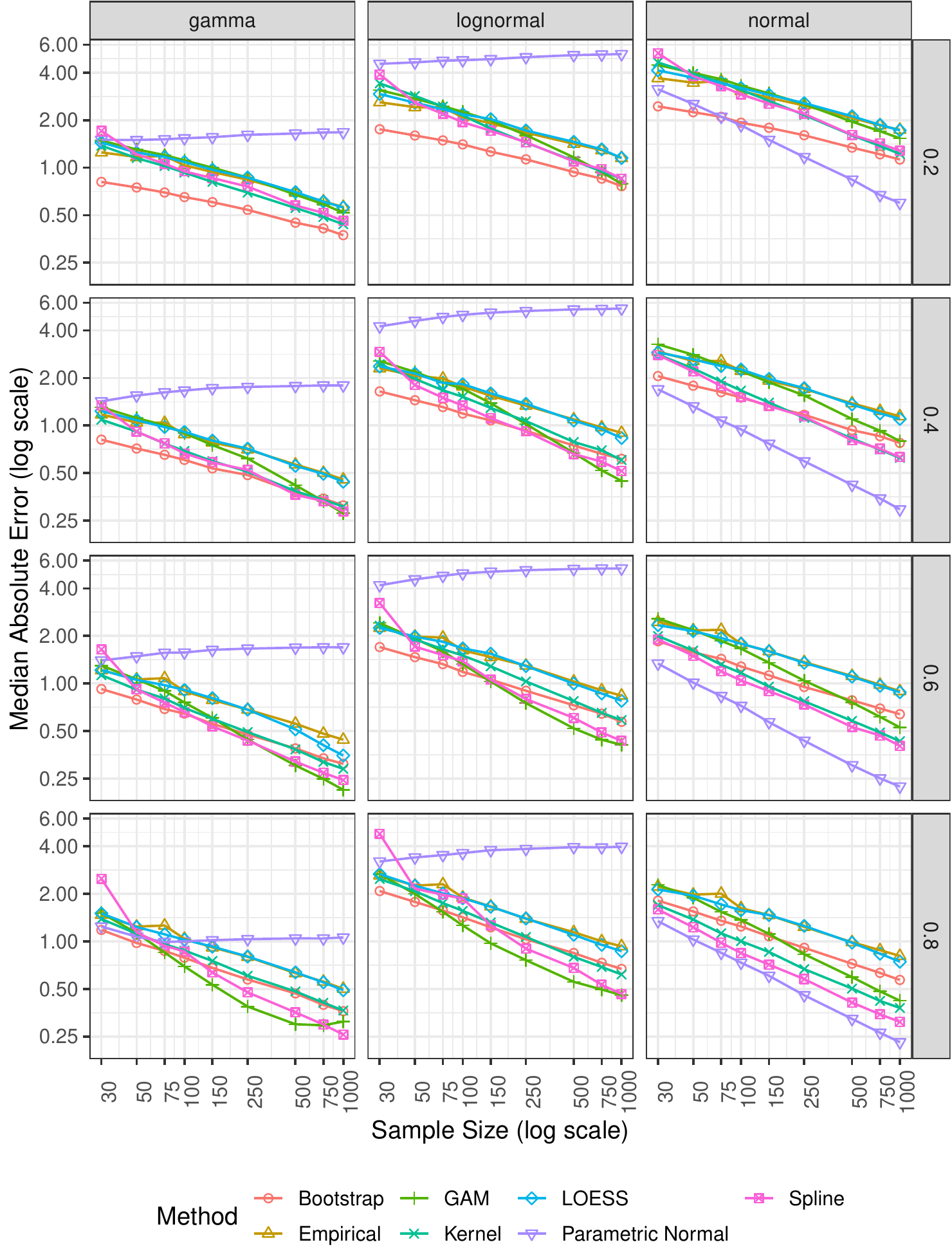} 

}

\caption[Comparative performance of cutpoint estimation methods on normal and lognormal data]{Comparative performance of cutpoint estimation methods on normal and lognormal data.}\label{fig:unnamed-chunk-2}
\end{figure}
\end{CodeChunk}

\hypertarget{performance-of-different-cutpoint-estimation-methods}{%
\section{Performance of different cutpoint estimation
methods}\label{performance-of-different-cutpoint-estimation-methods}}

In order to compare the performance of the different estimation methods,
we performed a simulation study similar to the one by Fluss et al.
\citep{fluss_estimation_2005}. In the present study we used the
aforementioned seven different methods (excluding the mean, median and
manual methods) to determine the optimal cutpoint for the Youden index.
Specifically, we drew samples from three types of distributions (normal,
log-normal, and gamma) and at four different levels of separation
(Youden index values of 0.2, 0.4, 0.6 and 0.8). In the case of normally
distributed data the target separation was achieved by keeping the mean
of the control group constant at 100 and manipulating the mean of the
experimental group (105.05, 110.49, 116.83 and 125.63). The standard
deviation for both groups was constant at 10. In the case of
log-normally distributed data the target separation was achieved by
keeping the mean of the control group constant at 2.5 and manipulating
the mean of the experimental group (2.76, 3.02, 3.34 and 3.78). The
standard deviation for both groups was constant at 2.5. In the case of
gamma distributed data the target separation was achieved by keeping the
rate parameter of the control group constant at 0.5 and manipulating the
rate parameter of the experimental group (0.344, 0.233, 0.143, 0.072).
The shape parameter for both groups was constant at 2. In all scenarios
the prevalence was held constant at 50 percent without loss of
generalizability \citep{leeflang_bias_2008}. The overall sample sizes
were 30, 50, 75, 100, 150, 250, 500, 750 and 1.000, resulting in 108
different scenarios. For each scenario 10.000 samples were generated. In
each sample we compared the optimal cutpoint identified by the different
methods to the true optimal cutpoint according to the scenario
parameters. To describe the performance, we calculated the median
absolute error within the scenarios. Figure 1 shows the performance of
the different methods.

Mirroring earlier results
\citep{fluss_estimation_2005, leeflang_bias_2008}, the most efficient
method to estimate optimal cutpoints depends on the underlying
distribution, sample size and level of separation. When the underlying
distributions are normal, the Normal method and for small samples
bootstrapping yield the best results. However, the Normal method gives
erroneous optimal cutpoints with lognormal and gamma distributions. The
empirical and LOESS methods show the highest levels of error when the
data is normally distributed. In that scenario, the level of error of
the empirical method with 1.000 observations is about as high as the
error of the Normal method with 30 observations. In scenarios with
non-normal distributions either bootstrapping or GAM give the best
results. Bootstrapping is better for small levels of separation and
small sample sizes while GAM outperforms all other methods for larger
levels of separation combined with larger sample sizes. While more work
is needed to tease apart the different conditions under which the
individual methods are optimal, it seems that more advanced methods are
preferable to the empirical method
\citep{fluss_estimation_2005, leeflang_bias_2008, hirschfeld_simulation_2014}.

\hypertarget{software}{%
\section{Software}\label{software}}

\hypertarget{overview-of-functions-classes-and-data}{%
\subsection{Overview of functions, classes and
data}\label{overview-of-functions-classes-and-data}}

The package consists of several main functions. Some of these functions
may serve as inputs to other functions or may also be used individually,
for example the cutpoint estimation methods and \texttt{roc}.

\begin{figure}
\centering
\includegraphics[width=\textwidth,height=0.4\textheight]{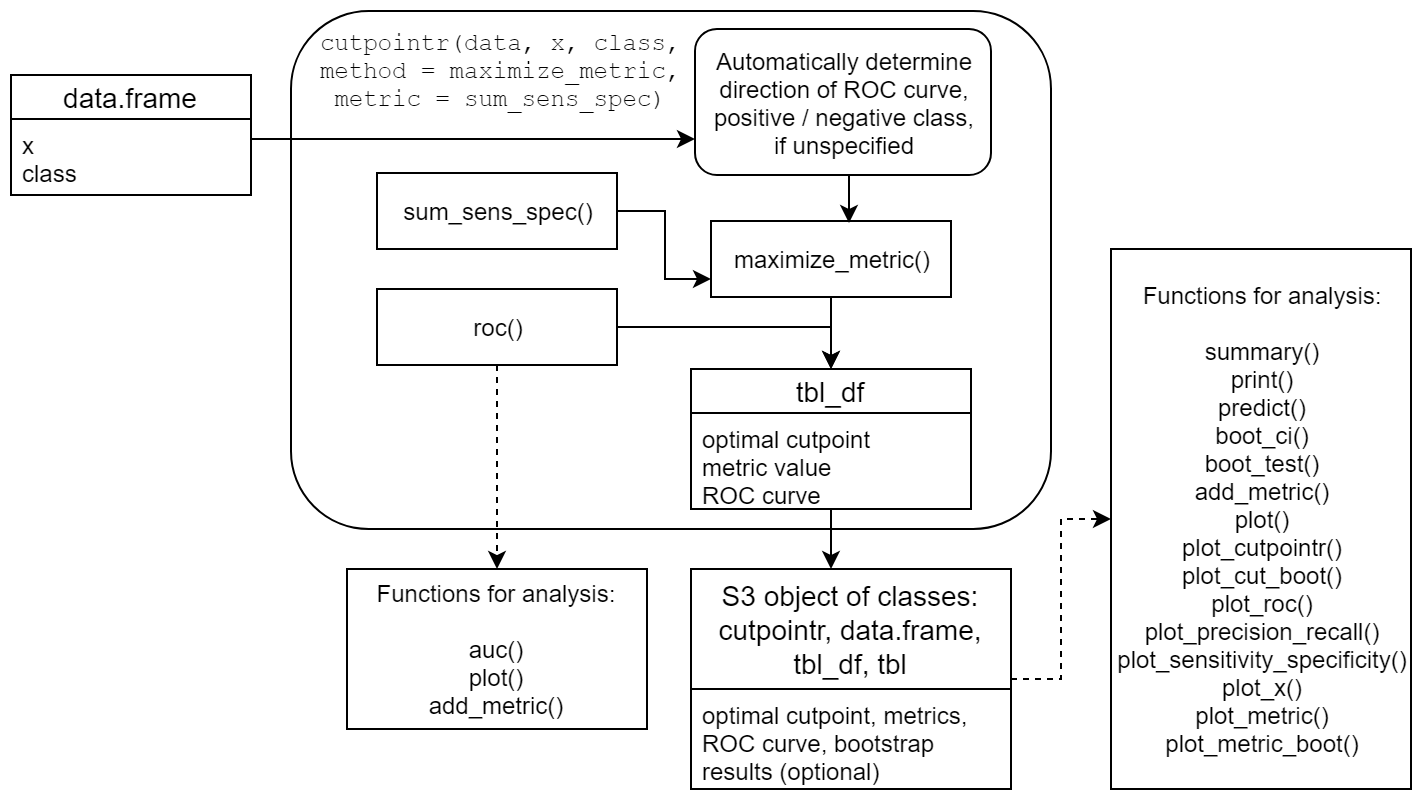}
\caption{Interplay of functions and classes in cutpointr.}
\end{figure}

\hypertarget{main-functions}{%
\subsubsection{Main functions}\label{main-functions}}

The main function of the package is \texttt{cutpointr}. There are two
ways to invoke \texttt{cutpointr}. First, the data can be supplied as a
data frame in the \texttt{data} argument and the predictor, outcome, and
grouping variables then must be given in \texttt{x}, \texttt{class}, and
\texttt{subgroup}. This interface uses nonstandard evaluation via the
\pkg{rlang} package. For programming purposes, supplied variables can be
unquoted using \texttt{!!}. Second, the \texttt{data} argument can be
left as \texttt{NULL} and the predictor, outcome, and grouping variables
then have to be supplied as raw vectors of any suitable type. Further
arguments control the choice of estimation method, which metric shall be
optimized, if bootstrapping should be run and more. Arguments to these
downstream functions are passed to \texttt{...} directly in
\texttt{cutpointr}. Some methods are prefixed with \texttt{oc\_} to make
them distinguishable from metrics, the maximization and minimization
methods are not. The functions for calculating metrics and for finding
cutpoints can easily be user-defined. For example, the function that is
supplied as the \texttt{metric} argument should return a numeric vector,
a matrix, or a data frame with one column. If the column is named, the
name will be included in the output and plots. The inputs are vectors of
\(TP\), \(FP\), \(TN\) and \(FN\). See \texttt{vignette("cutpointr")}
for more details on how to write functions that are suitable inputs for
\texttt{method} or \texttt{metric}. For a complete overview of metrics
included in \pkg{cutpointr} see table 2.

The \texttt{multi\_cutpointr} function is a wrapper that by default runs
\texttt{cutpointr} using all numeric columns in \texttt{data} as
predictors, except for the \texttt{class} column, or using the columns
in the \texttt{x} argument. All other arguments are simply passed to
\texttt{cutpointr}. The rows of the resulting data frame represent the
output of \texttt{cutpointr} per predictor variable. There is a
\texttt{summary} method for objects of class \texttt{multi\_cutpointr},
however the plotting functions do not support \texttt{multi\_cutpointr}
objects.

Both \texttt{cutpointr} and \texttt{multi\_cutpointr} return ROC curves.
If bootstrapping was run, also ROC curves based on the in- and
out-of-bag observations are returned within the \texttt{boot} column.
Internally, the \texttt{roc} function is used which is also exported so
that it can be used on its own if only a ROC curve is desired. It
returns the ROC curve as a \texttt{data.frame} with all unique cutpoints
and the associated numbers of correct and false classifications, the
associated rates and the obtained metric value in the column \texttt{m}
as calculated using the function in the \texttt{metric} argument. The
results can be inspected for example with \texttt{summary},
\texttt{plot} or \texttt{plot\_cutpointr}. The latter function allows
for flexibly plotting metrics per cutpoint or against each other,
including confidence intervals if bootstrapping was run.

\hypertarget{classes}{%
\subsubsection{Classes}\label{classes}}

The object returned by \texttt{cutpointr} is an object of the classes
\texttt{cutpointr}, \texttt{tbl\_df}, \texttt{tbl}, and
\texttt{data.frame}. It is returned and printed unaltered as a
\texttt{tbl\_df} to make sure that the output is comprehensible at a
glance, clearly arranged, and also easy to pass to subsequent functions.
In addition to the model parameters and some metrics, the object also
includes data frames in list columns that contain the original data and
the ROC curve. If a subgroup was defined, the aforementioned data are
split per subgroup and returned in the respective rows. The ROC curve is
returned as a nested data frame in the \texttt{roc\_curve} column with
the metric at each cutpoint \texttt{x.sorted} in the \texttt{m} column.
If bootstrapping was run, the \texttt{boot} column contains a data frame
with the bootstrap results. Each row represents one bootstrap
repetition. For each repetition, the AUC, the metric in \texttt{metric},
several other metrics, the numbers of true or false classifications and
the in- and out-of-bag ROC curves, again in nested data frames, are
returned. For all of these metrics \texttt{cutpointr} returns the in-
and out-of-bag value, as indicated by the suffix \texttt{\_b} or
\texttt{\_oob}. The output of the \texttt{summary} method for
\texttt{cutpointr} objects is an object of the classes
\texttt{summary\_cutpointr}, \texttt{tbl\_df}, \texttt{tbl} and
\texttt{data.frame}.

The object returned by \texttt{multi\_cutpointr} is an object of the
classes \texttt{multi\_cutpointr}, \texttt{tbl\_df}, \texttt{tbl}, and
\texttt{data.frame} that has a corresponding \texttt{summary} method,
but no plotting methods. The \texttt{roc} function also returns a
\texttt{tbl\_df} that carries an additional \texttt{roc\_cutpointr}
class.

\hypertarget{data}{%
\subsubsection{Data}\label{data}}

The data set \texttt{suicide} which is suitable for running binary
classification tasks is included in \pkg{cutpointr}. Various personality
and clinical psychological characteristics of 532 participants were
assessed as part of an online-study determining optimal cutpoints
\citep{glischinski_depressive_2016}. To identify persons at risk for
attempting suicide, various demographic and clinical characteristics
were assessed. Depressive Symptom Inventory - Suicidality Subscale
(DSI-SS) sum scores and whether the subject had previously attempted
suicide were recorded. Two additional demographic variables (age and
gender) are also included to test estimation of different cutpoints per
subgroup.

\hypertarget{example-suicide-dataset}{%
\section{Example: Suicide dataset}\label{example-suicide-dataset}}

To showcase the functionality, we'll use the included \texttt{suicide}
data set and calculate possible cutpoints for the DSI-SS. To estimate a
cutpoint maximizing \(Se + Sp\), various in-sample metrics, and the
ROC-curve \texttt{cutpointr} can be run with minimal inputs.

\begin{CodeChunk}

\begin{CodeInput}
R> library("cutpointr")
R> opt_cut <- cutpointr(suicide, dsi, suicide, silent = TRUE)
R> opt_cut
\end{CodeInput}

\begin{CodeOutput}
# A tibble: 1 x 16
  direction optimal_cutpoint method          sum_sens_spec      acc sensitivity
  <chr>                <dbl> <chr>                   <dbl>    <dbl>       <dbl>
1 >=                       2 maximize_metric       1.75179 0.864662    0.888889
  specificity      AUC pos_class neg_class prevalence outcome predictor
        <dbl>    <dbl> <fct>     <fct>          <dbl> <chr>   <chr>    
1    0.862903 0.923779 yes       no         0.0676692 suicide dsi      
            data roc_curve          boot 
  <list<df[,2]>> <list>             <lgl>
1      [532 x 2] <tibble [13 x 10]> NA   
\end{CodeOutput}
\end{CodeChunk}

The determined ``optimal'' cutpoint in this case is 2, thus all subjects
with a score of at least 2 would be classified as positives, indicating
a history of at least one suicide attempt. Based on this sample, this
leads to a sum of \(Se\) and \(Sp\) of 1.75. Furthermore, we receive
outputs that are independent of the determined cutpoint such as the AUC,
the prevalence, or the ROC curve.

\texttt{cutpointr} makes assumptions about the direction of the
dependency between \texttt{class} and \texttt{x} based on the median, if
\texttt{direction} and / or \texttt{pos\_class} and \texttt{neg\_class}
are not specified. Thus, the same result can be achieved by manually
defining \texttt{direction} and the positive / negative classes which is
slightly faster:

\begin{verbatim}
R> cutpointr(suicide, dsi, suicide, direction = ">=", pos_class = "yes",
+    neg_class = "no", method = maximize_metric, metric = sum_sens_spec)
\end{verbatim}

To receive additional descriptive statistics, \texttt{summary} can be
run on the \texttt{cutpointr} object (output omitted).

\hypertarget{reproducible-bootstrapping-and-parallelization}{%
\subsubsection{Reproducible Bootstrapping and
parallelization}\label{reproducible-bootstrapping-and-parallelization}}

If \texttt{boot\_runs} is larger than zero, \texttt{cutpointr} will
carry out the usual cutpoint calculation on the full sample, just as
before, and additionally on \texttt{boot\_runs} bootstrap samples. The
bootstrapping can be parallelized by setting
\texttt{allowParallel\ =\ TRUE} and starting a parallel backend.
Internally, the parallelization is carried out using \pkg{foreach}
\citep{Microsoft2017} and \pkg{doRNG} \citep{gaujoux_dorng:_2017} for
reproducible parallel loops. An example of a suitable parallel backend
is a SOCK cluster that can be started using the \pkg{snow} package
\citep{tierney_snow:_2009}. Reproducibility can be achieved by setting a
seed using \texttt{set.seed}, both in the case of sequential or parallel
execution.

\begin{verbatim}
R> library("doSNOW")
R> library("tidyverse")
R> cl <- makeCluster(2) # 2 cores
R> registerDoSNOW(cl)
R> set.seed(100)
R> opt_cut_b <- cutpointr(suicide, dsi, suicide, boot_runs = 1000,
+    silent = TRUE, allowParallel = TRUE)
R> stopCluster(cl)
R> opt_cut_b %>% select(optimal_cutpoint, data, boot)
\end{verbatim}

\begin{CodeChunk}

\begin{CodeOutput}
# A tibble: 1 x 3
  optimal_cutpoint           data boot                 
             <dbl> <list<df[,2]>> <list>               
1                2      [532 x 2] <tibble [1,000 x 23]>
\end{CodeOutput}
\end{CodeChunk}

The returned object includes a nested data frame in the column
\texttt{boot} that contains the optimal cutpoint per bootstrap sample
along with the metric calculated using the function in \texttt{metric}
and various additional metrics. The metrics are suffixed by \texttt{\_b}
to indicate in-bag results or \texttt{\_oob} to indicate out-of-bag
results. Now we receive a larger summary and additional plots upon
applying \texttt{summary} and \texttt{plot} (see figure 3).

\begin{CodeChunk}

\begin{CodeInput}
R> summary(opt_cut_b)
\end{CodeInput}
\end{CodeChunk}

\begin{verbatim}
Method: maximize_metric 
Predictor: dsi 
Outcome: suicide 
Direction: >= 
Nr. of bootstraps: 1000 

    AUC   n n_pos n_neg
 0.9238 532    36   496

 optimal_cutpoint sum_sens_spec    acc sensitivity specificity tp fn fp  tn
                2        1.7518 0.8647      0.8889      0.8629 32  4 68 428

Predictor summary: 
        Min.   5% 1st Qu. Median  Mean 3rd Qu.  95% Max.    SD NAs
overall    0 0.00       0      0 0.921       1 5.00   11 1.853   0
no         0 0.00       0      0 0.633       0 4.00   10 1.412   0
yes        0 0.75       4      5 4.889       6 9.25   11 2.550   0

Bootstrap summary: 
          Variable  Min.    5% 1st Qu. Median  Mean 3rd Qu.   95%  Max.    SD
  optimal_cutpoint 1.000 1.000   2.000  2.000 2.078   2.000 4.000 4.000 0.668
             AUC_b 0.832 0.878   0.907  0.926 0.923   0.941 0.960 0.981 0.025
           AUC_oob 0.796 0.867   0.901  0.923 0.924   0.955 0.974 0.990 0.034
   sum_sens_spec_b 1.552 1.661   1.722  1.760 1.757   1.794 1.837 1.925 0.054
 sum_sens_spec_oob 1.393 1.569   1.667  1.724 1.722   1.782 1.867 1.905 0.089
             acc_b 0.724 0.776   0.852  0.867 0.860   0.880 0.904 0.942 0.035
           acc_oob 0.704 0.763   0.845  0.862 0.855   0.879 0.900 0.928 0.038
     sensitivity_b 0.657 0.808   0.865  0.903 0.900   0.938 0.974 1.000 0.052
   sensitivity_oob 0.500 0.688   0.818  0.875 0.868   0.929 1.000 1.000 0.097
     specificity_b 0.709 0.762   0.849  0.864 0.857   0.878 0.908 0.951 0.039
   specificity_oob 0.691 0.751   0.845  0.861 0.854   0.879 0.911 0.948 0.044
    cohens_kappa_b 0.174 0.280   0.367  0.414 0.410   0.456 0.525 0.643 0.072
  cohens_kappa_oob 0.125 0.247   0.336  0.390 0.388   0.443 0.520 0.641 0.082
\end{verbatim}

\begin{CodeChunk}

\begin{CodeInput}
R> plot(opt_cut_b)
\end{CodeInput}
\begin{figure}

{\centering \includegraphics[width=1\linewidth]{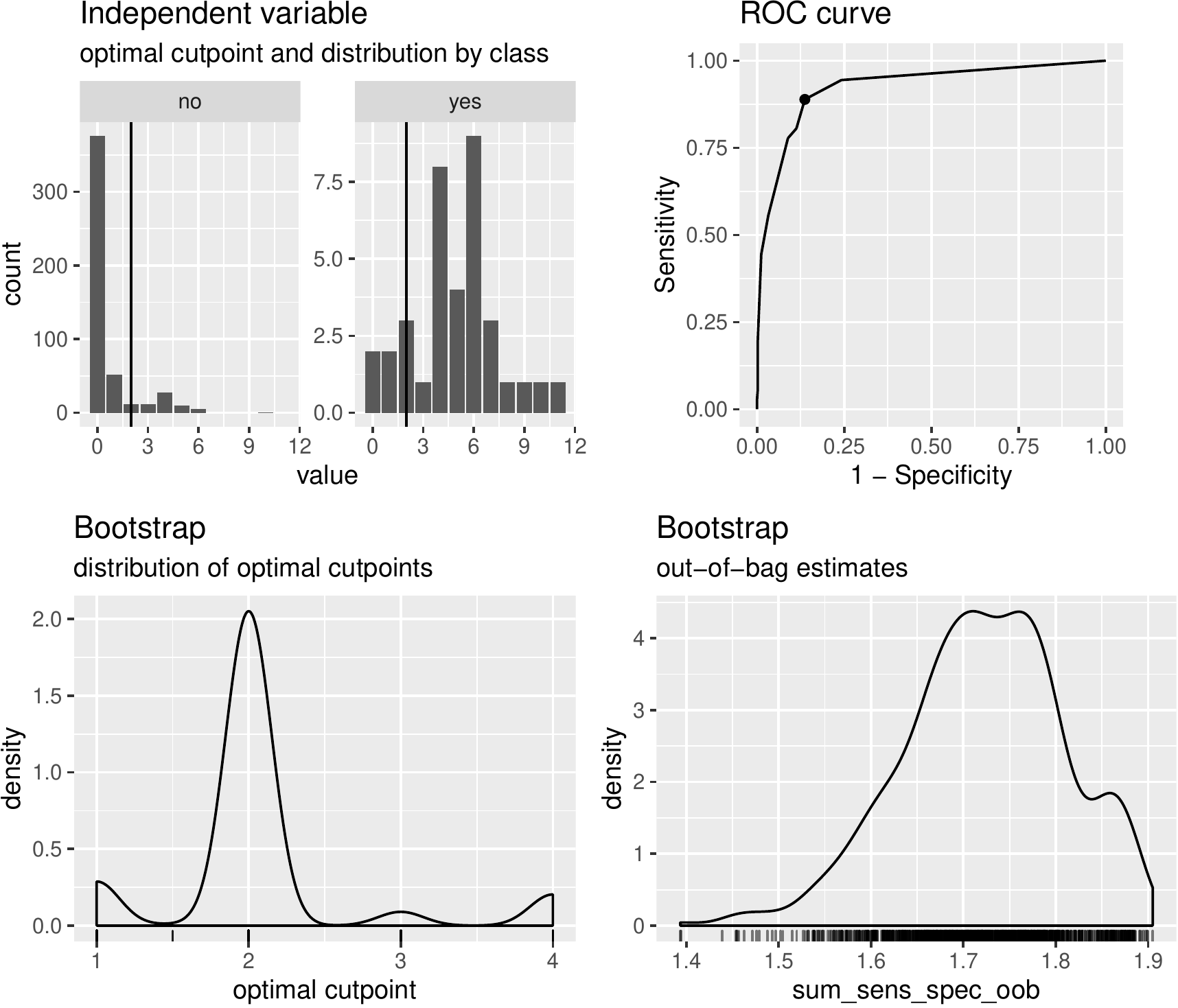} 

}

\caption[Main plotting function of cutpointr, the top left showing the distribution of the predictor values per class, the top right showing the ROC-curve, the bottom left showing the bootstrapped cutpoint variability, and the bottom right showing the distribution of the out-of-bag metric values]{Main plotting function of cutpointr, the top left showing the distribution of the predictor values per class, the top right showing the ROC-curve, the bottom left showing the bootstrapped cutpoint variability, and the bottom right showing the distribution of the out-of-bag metric values.}\label{fig:unnamed-chunk-6}
\end{figure}
\end{CodeChunk}

This allows us to form expectations about the performance on unseen data
and to compare these expectations to the in-sample performance. First,
the cutpoints that are obtained via the chosen method - maximizing the
sum of sensitivity and specificity - are quite stable here. In the
majority of bootstrap samples (777 out of 1000 times) the chosen
cutpoint is 2 as can be estimated from the plot or simply by inspecting
the results, for example using
\texttt{opt\_cut\_b\ \%\textgreater{}\%\ select(boot)\ \%\textgreater{}\%\ unnest(boot)\ \%\textgreater{}\%\ count(optimal\_cutpoint)}.
2 is also the cutpoint that was determined on the full sample. Second,
the distribution of the out-of-bag performance, namely \(Se + Sp\), can
be assessed from the plot. Summary statistics for the in- and out-of-bag
metrics are returned by the \texttt{summary} function, for example the
first and third quartiles of the out-of-bag values of \(Se + Sp\) are
1.67 and 1.78, respectively, which could also be calculated with
\texttt{boot\_ci(opt\_cut\_b,\ sum\_sens\_spec,\ in\_bag\ =\ FALSE,\ alpha\ =\ 0.5)}.
Furthermore, since both in- and out-of-bag results are returned, we can
confirm the optimism of the in-sample performance estimation. The median
of the in-bag metric is 1.76 while the median of the out-of-bag metric
is 1.72. The optimal cutpoint of 2 yields \(Se + Sp = 1.75\) on the full
sample. The other metrics can be compared in a similar fashion. Third,
to inspect the in-sample fit, the chosen cutpoint is displayed on a
histogram of the predictor per class and on the ROC-curve.

The individual plots can be reproduced using \texttt{plot\_x},
\texttt{plot\_roc}, \texttt{plot\_cut\_boot} and
\linebreak \texttt{plot\_metric\_boot}. Another way of inspecting the
selection of the optimal cutpoint is to plot all possible cutpoints
along with the corresponding metric values. This can be achieved using
the \texttt{plot\_metric} function which also plots bootstrapped
confidence intervals of the in-sample metric at each cutpoint, if
bootstrapping was run (see figure 4).

\begin{verbatim}
R> plot_metric(opt_cut_b, conf_lvl = 0.95) +
+    ylab("Sensitivity + Specificity") +
+    theme_bw()
\end{verbatim}

\begin{CodeChunk}
\begin{figure}

{\centering \includegraphics{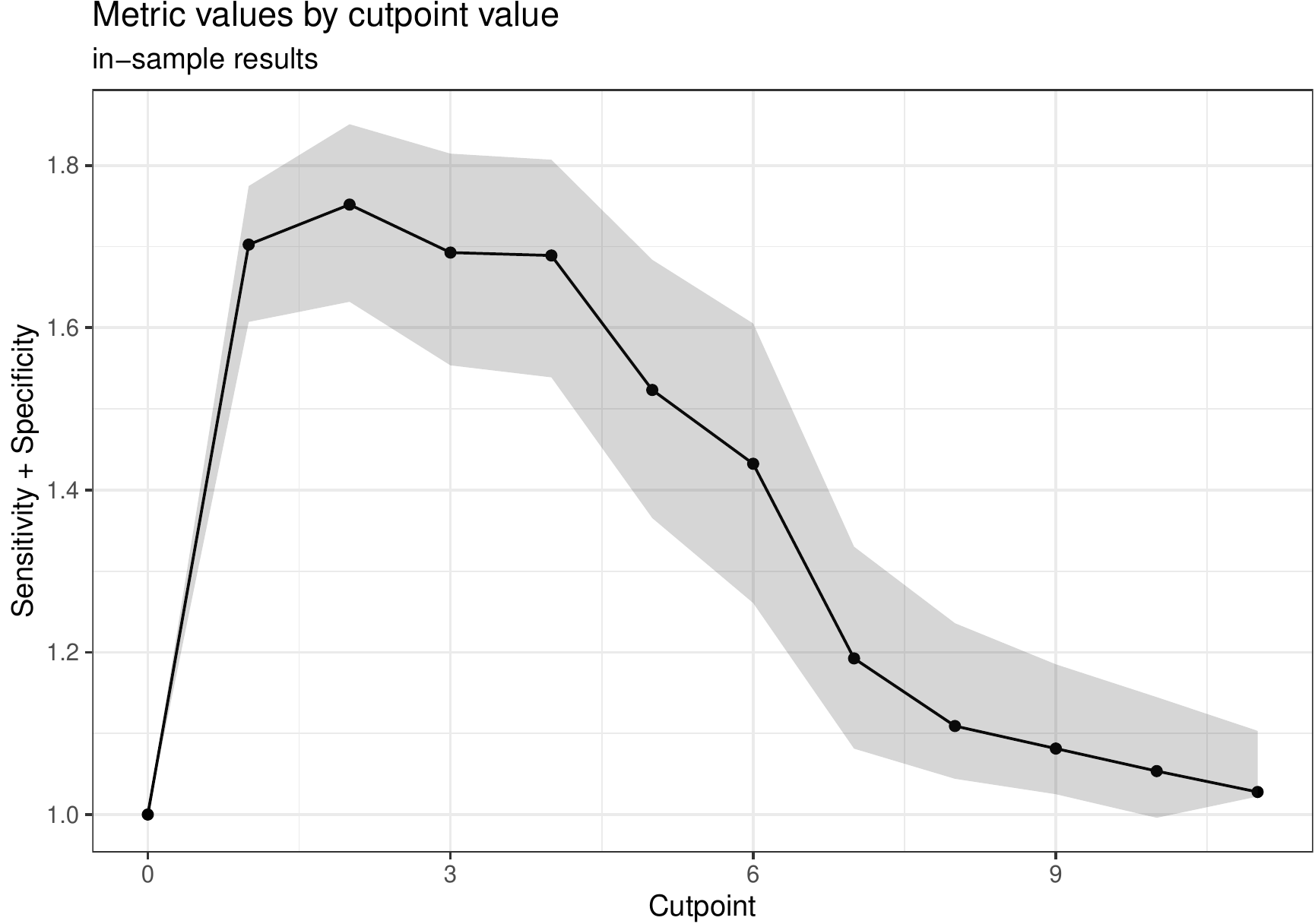} 

}

\caption[Values of the metric function per cutpoint with 95 percent bootstrap confidence interval]{Values of the metric function per cutpoint with 95 percent bootstrap confidence interval.}\label{fig:unnamed-chunk-7}
\end{figure}
\end{CodeChunk}

\hypertarget{subgroups}{%
\subsubsection{Subgroups}\label{subgroups}}

If the data contains a grouping variable that is assumed to play a role
in determining the cutpoint, the complete selection and validation
procedure can be conveniently run on separate subgroups by specifying
the \texttt{subgroup} argument.

\begin{verbatim}
R> set.seed(100)
R> opt_cut_b_g <- cutpointr(suicide, dsi, suicide, gender, boot_runs = 1000,
+    boot_stratify = TRUE, pos_class = "yes", direction = ">=")
R> opt_cut_b_g %>% select(subgroup, optimal_cutpoint, sum_sens_spec, data)
\end{verbatim}

\begin{CodeChunk}

\begin{CodeOutput}
# A tibble: 2 x 4
  subgroup optimal_cutpoint sum_sens_spec           data
  <chr>               <dbl>         <dbl> <list<df[,2]>>
1 female                  2       1.80812      [392 x 2]
2 male                    3       1.62511      [140 x 2]
\end{CodeOutput}
\end{CodeChunk}

This time the output contains one row per subgroup. In this case, we
receive different optimal cutpoints for the female and male group. As we
can also see from the in-sample results, the performance seems to be
superior in the female group. To confirm this, we can again inspect the
summary, particularly the bootstrap results.

\begin{CodeChunk}

\begin{CodeInput}
R> summary(opt_cut_b_g)
\end{CodeInput}
\end{CodeChunk}

\begin{verbatim}
Method: maximize_metric 
Predictor: dsi 
Outcome: suicide 
Direction: >= 
Subgroups: female, male 
Nr. of bootstraps: 1000 

Subgroup: female 
--------------------------------------
[...]
Bootstrap summary:
          Variable  Min.    5% 1st Qu. Median  Mean 3rd Qu.   95%  Max.    SD
[...]
 sum_sens_spec_oob 1.338 1.609   1.729  1.783 1.779   1.864 1.903 1.945 0.095
[...]

Subgroup: male 
-------------------------------------
[...]
Bootstrap summary:
          Variable  Min.    5% 1st Qu. Median  Mean 3rd Qu.   95%  Max.    SD
[...]
 sum_sens_spec_oob 0.786 0.981   1.332  1.504 1.479   1.659 1.860 2.000 0.248
[...]
\end{verbatim}

The out-of-bag value of the mean of the sum of sensitivity and
specificity in the female group is about 0.29 larger than in the male
group. Additionally, the selected cutpoints by simple maximization are
less variable in the female group (see also figure 5).

\begin{CodeChunk}

\begin{CodeInput}
R> plot(opt_cut_b_g)
\end{CodeInput}
\begin{figure}

{\centering \includegraphics{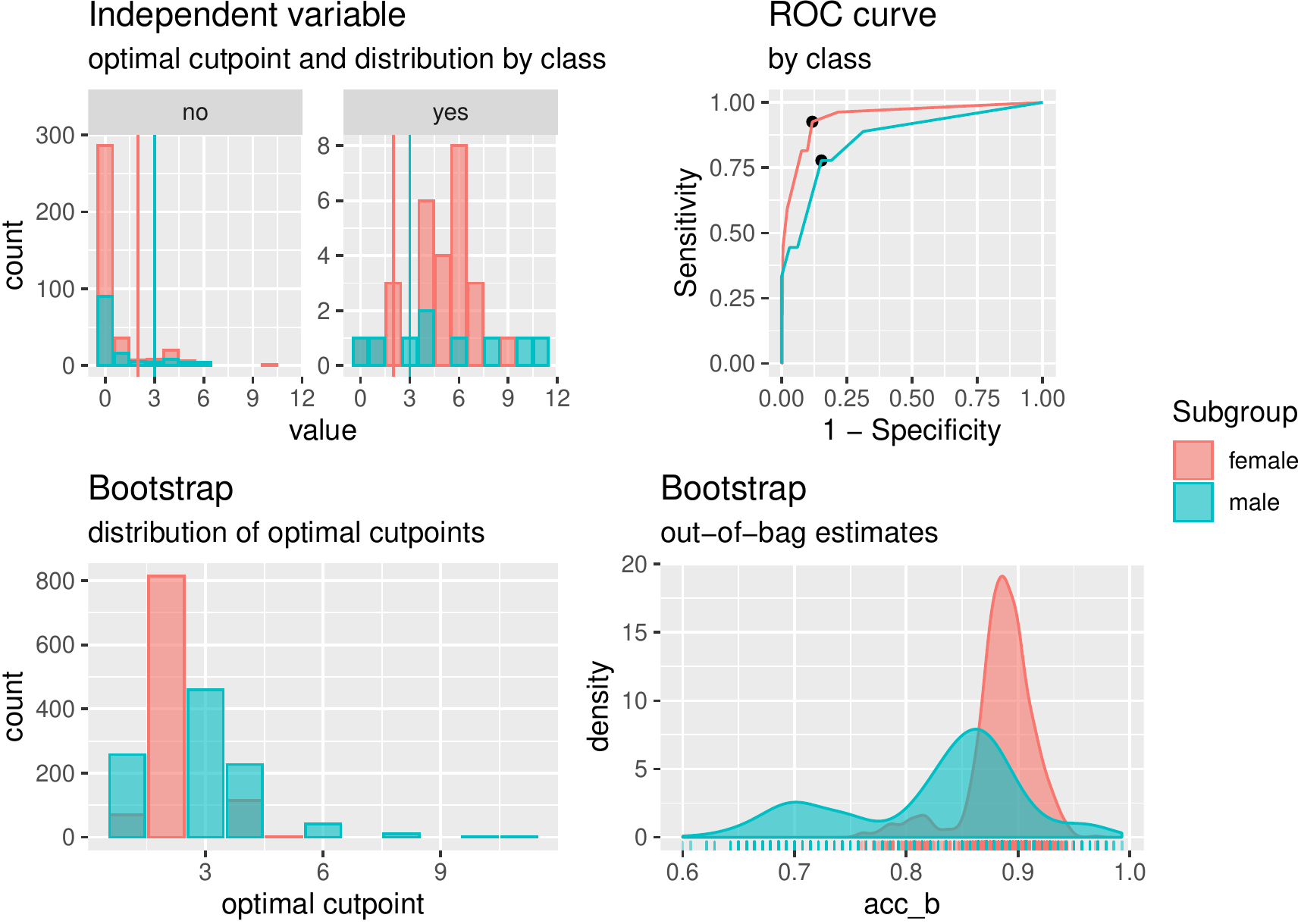} 

}

\caption[Main plotting function of cutpointr with subgroups showing the distribution of the predictor values per class, the ROC-curve, the bootstrapped cutpoint variability, and the distribution of the out-of-bag metric values]{Main plotting function of cutpointr with subgroups showing the distribution of the predictor values per class, the ROC-curve, the bootstrapped cutpoint variability, and the distribution of the out-of-bag metric values.}\label{fig:unnamed-chunk-10}
\end{figure}
\end{CodeChunk}

The higher predictive performance of the predictor variable in the
female group is furthermore emphasized by the larger AUC. The summary of
\texttt{AUC\_b} summarizes the AUC in all bootstrap samples which can be
informative for comparing the general predictive ability of the
predictor variable in all subgroups.

The observed variability of the selected cutpoints showcases the need
for methods that estimate optimal cutpoints with less variability.
\pkg{cutpointr} offers \texttt{maximize\_boot\_metric},
\linebreak \texttt{minimize\_boot\_metric},
\texttt{maximize\_gam\_metric}, \texttt{minimize\_gam\_metric},
\linebreak \texttt{maximize\_spline\_metric},
\texttt{minimize\_spline\_metric}, \texttt{maximize\_loess\_metric},
\linebreak \texttt{minimize\_loess\_metric}, \texttt{oc\_youden\_normal}
and \texttt{oc\_youden\_kernel} for that purpose. Since we can not
assume the predictor in the positive and negative class to be normally
distributed, we will use the Kernel method for determining optimal
cutpoints to maximize the Youden index. In our experience, the Kernel
method is more suitable for data with a small number of unique predictor
values than the smoothing methods. The function
\texttt{oc\_youden\_kernel} does not return a metric value at the
estimated cutpoint on its own, but instead the estimated cutpoint is
applied to the ROC curve, see chapter 2.2. The function in
\texttt{metric} is only used for validation purposes so that we can
compare these results to the previous ones.

\newpage

\begin{verbatim}
R> set.seed(100)
R> opt_cut_k_b_g <- cutpointr(suicide, dsi, suicide, gender,
+    method = oc_youden_kernel, metric = sum_sens_spec, boot_runs = 1000)
R> summary(opt_cut_k_b_g)

Method: oc_youden_kernel 
Predictor: dsi 
Outcome: suicide 
Direction: >= 
Subgroups: female, male 
Nr. of bootstraps: 1000 

Subgroup: female 
--------------------------------------
[...]
Bootstrap summary:
          Variable  Min.    5% 1st Qu. Median  Mean 3rd Qu.   95%  Max.    SD
[...]
 sum_sens_spec_oob 1.338 1.590   1.741  1.792 1.785   1.873 1.906 1.945 0.102
[...]

Subgroup: male 
-------------------------------------
[...]
Bootstrap summary: 

          Variable  Min.    5% 1st Qu. Median  Mean 3rd Qu.   95%  Max.    SD
[...]
 sum_sens_spec_oob 0.732 0.975   1.378  1.542 1.537   1.787 1.857 2.000 0.263
[...]
\end{verbatim}

The plot of \texttt{opt\_cut\_k\_b\_g} (not shown) reveals that the
determined cutpoints are again much less variable in the female than in
the male group. Since the Kernel method should exhibit a lower
variability than the previously used maximization method, we expect
better validation results in the bootstrapping. Indeed, in both
subgroups \(Se + Sp\) in the out-of-bag observations is slightly larger
than before with the nonparametric empirical method (for example, the
mean of \(Se + Sp\) in the male group is 1.54 here instead of 1.48).

\hypertarget{alternative-metric-function}{%
\subsubsection{Alternative metric
function}\label{alternative-metric-function}}

Various alternative metric functions are available and in the example
dataset it is reasonable to assign different costs to false positives
and false negatives. A false positive may lead to someone being treated
without an urgent need to do so, while a false negative may lead to
someone not being treated who may attempt suicide or suffer from a
considerably lower quality of life. \pkg{cutpointr} offers the
possibility to define these separate costs using the
\texttt{misclassification\_cost} metric. Since we want to minimize the
costs we use the \texttt{minimize\_metric} function as \texttt{method}.
If also separate utilities for true positives and true negatives should
be defined, we could use the \texttt{total\_utility} function. In
practice, the ratio of the costs or the actual costs may be known
exactly but for the purpose of this example we assume that a false
negative is ten times more costly than a false positive. Incidentally,
this leads to cutpoints that are comparable to the previously obtained
ones.

\begin{verbatim}
R> set.seed(100)
R> opt_cut_c_g_b <- cutpointr(suicide, dsi, suicide, gender,
+    method = minimize_metric, metric = misclassification_cost, cost_fp = 1,
+    cost_fn = 10, pos_class = "yes", direction = ">=", boot_runs = 1000)
R> opt_cut_c_g_b %>%
+    select(subgroup, optimal_cutpoint, misclassification_cost)
\end{verbatim}

\begin{CodeChunk}

\begin{CodeOutput}
# A tibble: 2 x 3
  subgroup optimal_cutpoint misclassification_cost
  <chr>               <dbl>                  <dbl>
1 female                  2                     63
2 male                    3                     40
\end{CodeOutput}
\end{CodeChunk}

\hypertarget{midpoints-as-optimal-cutpoints}{%
\subsection{Midpoints as optimal
cutpoints}\label{midpoints-as-optimal-cutpoints}}

So far - which is the default in \texttt{cutpointr} - we have considered
all unique values of the predictor as possible cutpoints. An alternative
could be to use a sequence of equidistant values instead, for example in
the preceding application all integers in \([0, 12]\) and the additional
cutpoint \(Inf\) to allow for classifying all observations as negative.
However, in the case of very sparse data and small intervals between the
defined candidate cutpoints (i.e., a ``dense'' sequence like
\texttt{seq(0,\ 12,\ by\ =\ 0.01)}) this leads to the uninformative
evaluation of large ranges of cutpoints that all result in the same
metric value. A more elegant alternative, not only for the case of
sparse data, that is supported by \pkg{cutpointr} is the use of a mean
value of the optimal cutpoint and the next highest (if
\texttt{direction\ =\ "\textgreater{}="}) or the next lowest (if
\texttt{direction\ =\ "\textless{}="}) predictor value in the data. The
result is an optimal cutpoint that is equal to the cutpoint that would
be obtained using an infinitely dense sequence of candidate cutpoints
and is thus more computationally efficient. This behavior can be
activated by setting \texttt{use\_midpoints\ =\ TRUE}. If we use this
setting, we obtain an optimal cutpoint of 1.5 for the complete sample on
the \texttt{suicide} data instead of 2 when maximizing \(Se + Sp\). In
practice, this would not make a difference here as the DSI-SS takes on
only integer values.

Furthermore, not using midpoints but the values found in the data may
lead to a positive or negative bias of the estimation procedure,
depending on \texttt{direction}. For example, if
\texttt{direction\ =\ "\textgreater{}="} and
\texttt{use\_midpoints\ =\ FALSE}, the returned optimal cutpoint
represents the highest possible cutpoint that leads to the optimal
metric, because all values that are below the optimal cutpoint and are
larger than the next lowest value in the data result in the same metric
value. This bias only applies to estimation methods that use the ROC
curve for selecting a cutpoint and are mainly relevant with small and
sparse data. See \texttt{vignette("cutpointr")} for more details and
some simulation results regarding the bias and its reduction using
midpoints.

\newpage

\hypertarget{benchmarks}{%
\subsection{Benchmarks}\label{benchmarks}}

To offer a comparison to established solutions, \pkg{cutpointr} 1.0.0
was benchmarked against \texttt{optimal.cutpoints} from
\pkg{OptimalCutpoints} 1.1-4 \citep{lopez-raton_optimalcutpoints:_2014},
\texttt{thres2} from \pkg{ThresholdROC} 2.7
\citep{perez-jaume_thresholdroc:_2017}, a custom function using the
functions \texttt{prediction} and \texttt{performance} from \pkg{ROCR}
1.0-7 \citep{sing_rocr:_2005} and a custom function using \texttt{roc}
from \pkg{pROC} v1.15.0 \citep{robin_proc:_2011} with the setting
\texttt{algo\ =\ 2} which is the fastest one in this application. As the
results will be identical for empirical maximization and more advanced
methods are not included in most other packages, the benchmark focuses
on a timing comparison. By generating data sets of different sizes, the
benchmarks offer a comparison of the scalability of the different
solutions and their performance on small and larger data.

The benchmarking was carried out unparallelised on Windows 10
v10.0.17763 Build 17763 with a 6-Core AMD CPU at 4GHz and 32GB of memory
running R 3.6.0. The runtime of the functions was assessed using
\pkg{microbenchmark} \citep{Mersmann2018}. The values of the predictor
variable were drawn from a normal distribution, leading to a number of
possible cutpoints that is close or equal to the sample size.
Accordingly, the search for an optimal cutpoint is relatively demanding.

Benchmarks are run for sample sizes of 100, 1000, 10000, \(1e5\),
\(1e6\), and \(1e7\). In small samples \pkg{cutpointr} is slower than
the other packages. This disadvantage seems to be caused mainly by
certain functions from \pkg{tidyr} that are used to nest the input data
and produce the compact output. The \proglang{C++} functions in
\pkg{cutpointr} lead to a sizeable speedup with larger data, but some
are slightly slower with small data. While the speed disadvantage with
small data should be of low practical importance, since it usually
amounts to only minimal delays, \pkg{cutpointr} scales more favorably
than the other solutions and is the fastest solution for sample sizes
\(\geq 1e6\). For data as large as \(1e7\) observations \pkg{cutpointr}
takes about 70\% of the runtime of \pkg{ROCR} and \pkg{pROC}. This
phenomenon is illustrated also by the performance in the task of ROC
curve calculation, in which \texttt{roc} from \pkg{cutpointr} is fast
also in small samples but has an advantage over \pkg{ROCR} and
\pkg{pROC} again only in larger samples (see figure 6).

All of \pkg{cutpointr}, \pkg{pROC} and \pkg{ROCR} are generally faster
than \pkg{OptimalCutpoints} and \pkg{ThresholdROC} with the exception of
small samples. \pkg{OptimalCutpoints} and \pkg{ThresholdROC} had to be
excluded from benchmarks with more than 10000 observations due to high
memory requirements. Upon inspection of the source code of the different
packages, it seems that the differences in speed and memory consumption
are mainly due to the way the ROC curve is built or whether a ROC curve
is built at all. A search over the complete vector of predictor values
is obviously inefficient, so a search over the points on the ROC curve
(which represent combinations of \(Se\) and \(Sp\) for all unique
predictor values) will offer a faster solution. For building the ROC
curve, a binary vector representing whether an observation is a \(TP\)
can be sorted according to the order of the predictor variable and
cumulatively summed up. The solutions by \pkg{cutpointr} and \pkg{ROCR}
rely on this algorithm.

\begin{CodeChunk}
\begin{figure}

{\centering \includegraphics{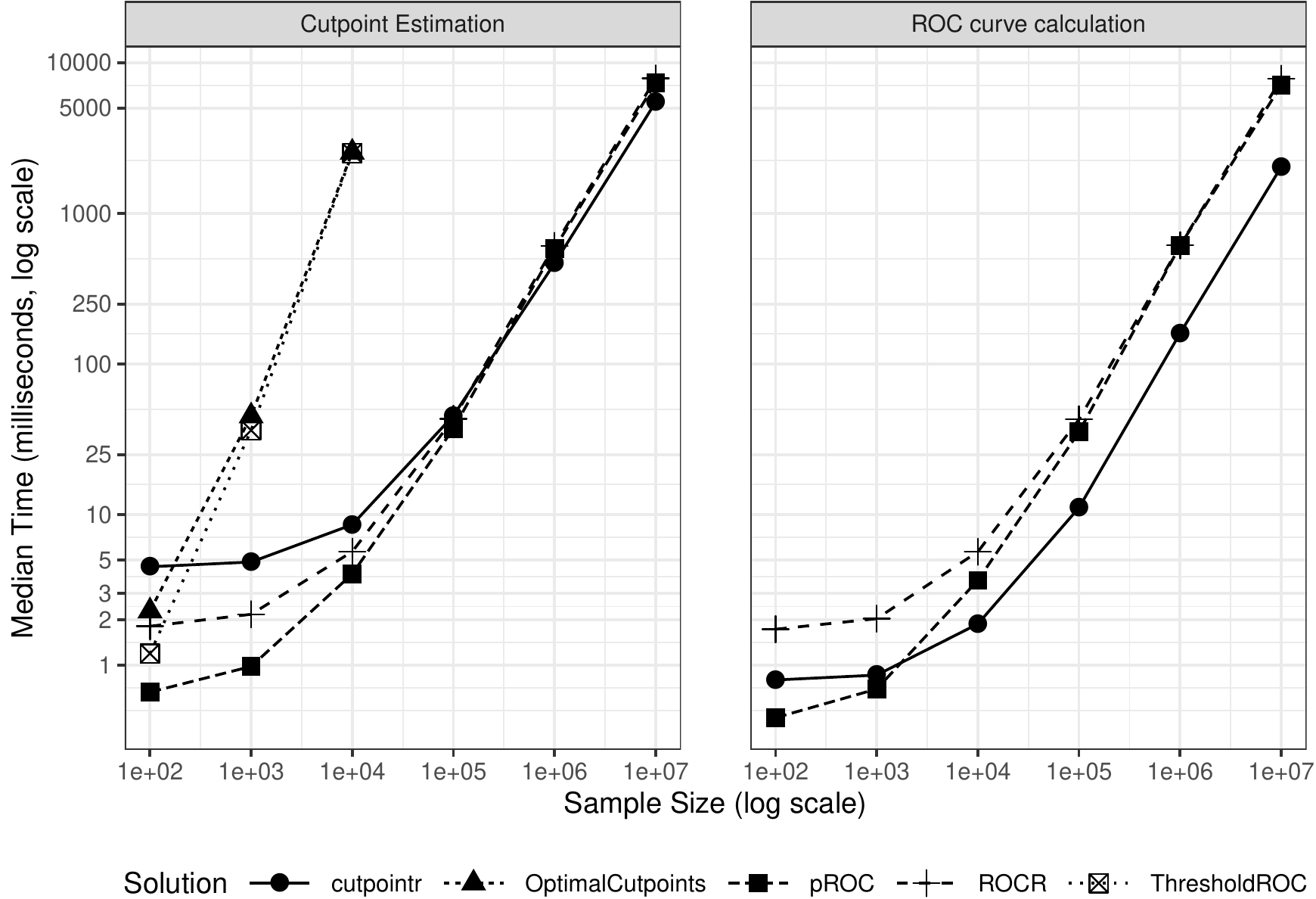} 

}

\caption[Benchmark results]{Benchmark results.}\label{fig:unnamed-chunk-14}
\end{figure}
\end{CodeChunk}

\hypertarget{conclusion}{%
\section{Conclusion}\label{conclusion}}

The aim in developing \pkg{cutpointr} was to make improved methods to
estimate and validate optimal cutpoints available. While there are
already several packages that aid the calculation of a wide range of
metrics
\citep{lopez-raton_optimalcutpoints:_2014, lopez-raton_gsympoint:_2017, perez-jaume_thresholdroc:_2017},
some of these rely on suboptimal procedures to estimate a cutpoint. Most
use the empirical method, which is prone to high variability an to
overestimating the diagnostic utility of optimal cutpoints.
\pkg{cutpointr} incorporates several methods to choose cutpoints that
are not necessarily optimal in the specific sample but perform
relatively well in the population from which the sample was drawn and
are more stable.

It is beyond the scope of the present manuscript to determine the best
estimation method, since this will likely depend on characteristics of
the study as well as the chosen metric. However, we believe that
\pkg{cutpointr} may prove useful for simulation studies because of its
scalability as well as applied research because of its ability to
estimate the cutpoint variability and out-of-sample performance.

\hypertarget{acknowledgements}{%
\section{Acknowledgements}\label{acknowledgements}}

The study was supported by the German Federal Ministry for Education and
Research to GH (BMBF \#01EK1501).

\renewcommand\refname{References}
\bibliography{Samplesizeyouden.bib}

\end{document}